\documentclass[runningheads]{llncs}
\usepackage{graphicx}
\usepackage{todonotes}
\usepackage{bbding}
\usepackage{booktabs}
\usepackage{multirow,bigdelim}
\usepackage{color}
\usepackage{xr}

\usepackage{amsmath,amsfonts,amssymb}
\newcommand{\bftab}{\fontseries{b}\selectfont}
\usepackage{rotating}
\usepackage{mwe}
\usepackage{graphbox}
\usepackage{multirow}
\begin{document}
\title{Unsupervised Cross-Modality Domain Adaptation for Segmenting Vestibular Schwannoma and Cochlea with Data Augmentation and Model Ensemble}

\titlerunning{Unsupervised Domain Adaptation for Segmenting VS and Cochlea}
%
%
\author{Hao Li\inst{1} \and
Dewei Hu\inst{1} \and
Qibang Zhu\inst{2} \and
Kathleen E.\ Larson\inst{3} \and
Huahong Zhang\inst{2} \and
Ipek Oguz\inst{2}
}
\authorrunning{H. Li et al.}
%
\institute{Department of Electrical and Computer Engineering \and
Department of Computer Science \and
Department of Biomedical Engineering \\ Vanderbilt University, Nashville, TN 37235, USA
}
%
\maketitle              
\begin{abstract}

Magnetic resonance images (MRIs) are widely used to quantify the volume of the vestibular schwannoma (VS) and cochlea. Recently, deep learning methods have shown state-of-the-art performance for segmenting these structures. However, training segmentation models may require manual labels in target domain, which is expensive and time-consuming. To overcome this problem, unsupervised domain adaptation is an effective way to leverage information from source domain to obtain accurate segmentations without requiring manual labels in target domain. In this paper, we propose an unsupervised learning framework to segment the VS and cochlea. Our framework leverages information from contrast-enhanced T1-weighted (ceT1-weighted) MRIs and its labels, and produces segmentations for T2-weighted MRIs without any labels in the target domain. We first applied a generator to achieve image-to-image translation. Next, we combined outputs from an ensemble of different models to obtain final segmentations. To cope with MRIs from different sites/scanners, we applied various ‘online’ data augmentations during training to better capture the geometric variability and the variability in image appearance and quality. Our method is easy to build and produces promising segmentations, with a mean Dice score of 0.7930 and 0.7432 for VS and cochlea respectively in the validation set of the cross-MoDA challenge.

\keywords{Unsupervised Domain adaptation \and Segmentation \and Vestibular schwannoma \and Cochlea \and Deep learning.}
\end{abstract}
\section{Introduction}
Vestibular schwannoma (VS) is a benign tumor of the human hearing system. For better understanding the disease progression, quantitative analysis of VS and cochlea from magnetic resonance images (MRIs) is important. Recently, deep learning frameworks have been dominating the medical segmentation field \cite{shapey2019artificial,wang2019automatic,dorent2020scribble,li2021mri,Shapey:SDATA:2021,ronneberger2015u} with state-of-the-art performances. However, supervised learning methods often require a high level of consistency between training and testing data. Consequently, such supervised methods often lack domain generalizability or ability to deal with images from various sites that have different intensity distributions, i.e., distribution shift or domain shift. Such a shift is usually caused by different image acquisition protocols or scanners; different image modalities could also be considered a domain shift problem. 

Furthermore, in medical image analysis, lack of human delineations in one or multiple domains is another common issue, which is problematic for supervised learning. Unsupervised domain adaptation (UDA) is a solution for increasing generalizability of deep learning models to deal with new data from different domains.

In the cross-MoDA challenge\url{\footnote{https://crossmoda.grand-challenge.org/}}, the ceT1-weighted and T2-weighted MRIs are provided, but only ceT1-weighted MRIs are labeled by experts. To obtain the segmentations on T2-weighted MRIs, we consider it as a UDA problem and propose an unsupervised cross-modality domain adaptation framework for segmenting the VS and cochlea. Our framework contains 2 parts: synthesis and segmentation. For synthesis, we apply a CycleGAN \cite{zhu2017unpaired} to perform unpaired image translation between ceT1-weighted and T2-weighted MRIs. For segmentation, we use the generated T2-weighted MRIs as input and train an ensemble of models with various data augmentations, each of which yields candidate segmentations of VS and cochlea. We fuse those candidate segmentations to form the final segmentation.

\section{Related work}
Supervised learning is an effective way for medical image segmentation when sufficient labels are available in target domain. Wang et al.~\cite{wang2019automatic} proposed an attention-based 2.5D convolutional neural network (CNN) to segment VS from T2-weighted MRIs with anisotropic resolution. In a following publication, Shapey et al.~\cite{shapey2019artificial} employed this 2.5D CNN and further explored the performance of segmenting VS on both T1-weighted and T2-weighted MRIs. As we know, obtaining manual annotations is labor-intensive and time-consuming. Dorent et al.~\cite{dorent2020scribble} introduced a novel weakly-supervised domain adaptation framework for VS segmentation on T2-weighted MRIs. In their work, only scribbles are needed as weak supervision in the target domain. They leveraged information from T1-weighted (source domain) MRIs to segment VS in the target domain based on co-segmentation and structured learning. However, in scenarios where there is no label available in target domain, UDA is a solution. Typical UDA methods try to align the image features between source domain and target domain  \cite{huo2018adversarial,chen2019unsupervised}. Once the features are well aligned, downstream tasks, such as segmentation, are relatively easy to accomplish. CycleGAN \cite{zhu2017unpaired} and MUNIT \cite{huang2018multimodal} are popular methods to achieve unpaired image translation. Huo et al.~\cite{huo2018adversarial} propose an end-to-end framework for unpaired synthesis between CT and MRI images, and jointly segment the spleen on CT images without any label from CT images during training. In their framework \cite{huo2018adversarial}, CycleGAN is used for unpaired image translation. Chen et al.~\cite{chen2019unsupervised} present a method to segment the cardiac structures from late-gadolinium enhanced (LGE) images by leveraging information from balanced steady-state free precession (bSSFP) images. Similarly, no label is used from target domain (LGE images) during training. The MUNIT is used as image translation network. While our framework is similar to the approach proposed by Huo et al.~\cite{huo2018adversarial}, we use an ensemble model with various augmentation strategies for our segmentation component to improve the robustness of our results. 

\section{Methods and material}
\subsection{Dataset}
The cross-modality domain adaptation for medical image segmentation challenge dataset (cross-MoDA\url{\footnote{https://crossmoda.grand-challenge.org/CrossMoDA/}}) contains two different MRI modalities: contrast-enhanced T1-weighted (ceT1-weighted) with an in-plane resolution of $0.4\times0.4 \mathrm{mm}$ and slice thickness between $1-1.5 \mathrm{mm}$, and high-resolution T2-weighted with an in-plane resolution of $0.5\times0.5 \mathrm{mm}$ and slice thickness between $1-1.5 \mathrm{mm}$. ceT1-weighted imaging was performed with an MPRAGE sequence and T2-weighted imaging with 3D CISS or FIESTA sequence. The training set contains 105 ceT1-weighted and 105 T2-weighted MRIs, and the validation set contains 32 T2-weighted MRIs. Expert manual VS and cochlea labels are available for the 105 ceT1-weighted training MRIs. More detailed information about this dataset can be found in \url{\footnote{https://wiki.cancerimagingarchive.net/pages/viewpage.action?pageId=70229053}}. 

\begin{figure}[t]
\centering
\includegraphics[width=1\linewidth]{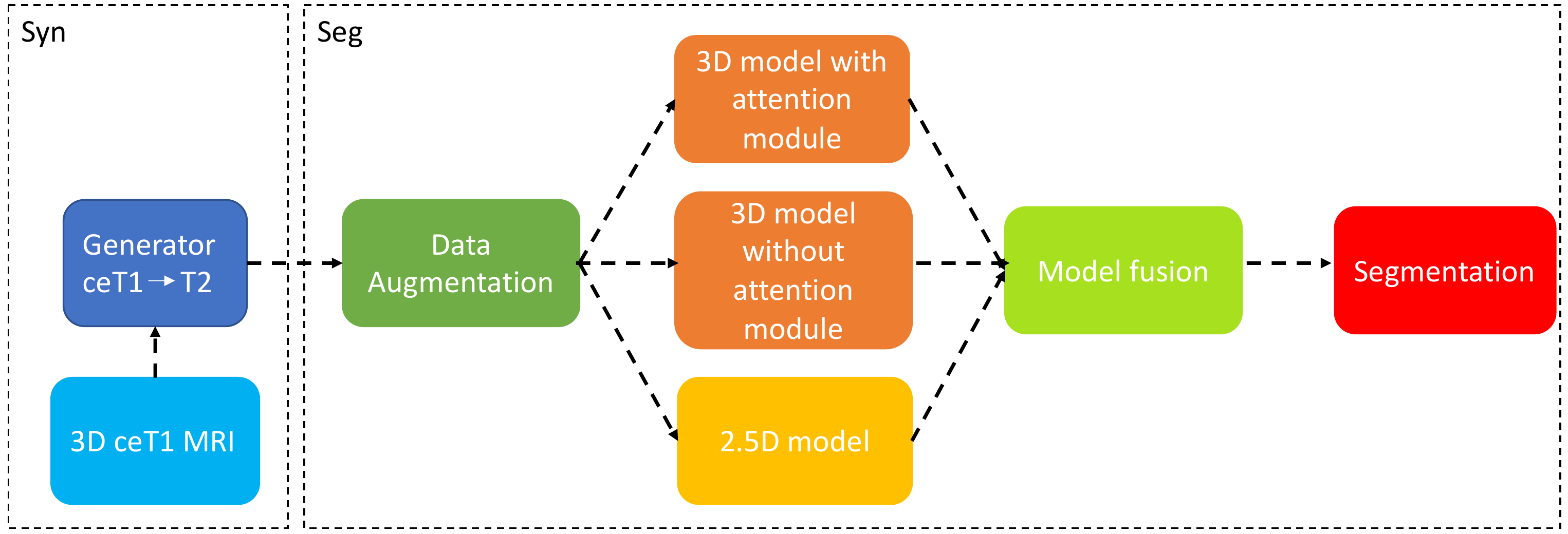}
\caption{Proposed overall framework. Our framework contains 2 parts: synthesis (Syn) and segmentation (Seg). We use a CycleGAN model as generator in the synthesis part. Results from the different models (and with different augmentations) are fused to obtain the final segmentation.
}
\label{network}
\end{figure}


\subsection{Overall framework}
Fig.~\ref{network} displays our proposed framework. There are two parts in our framework: synthesis (Syn) and segmentation (Seg). For the Syn part, the 3D ceT1-weighted image, after undergoing pre-processing, is fed to the CycleGAN \cite{zhu2017unpaired} pipeline to achieve image-to-image translation (ceT1-weighted to T2-weighted). Data augmentation strategies are applied on the generated T2-weighted MRIs to increase the model robustness. We send the augmented data into four different models: a 2.5D model with two different augmentation schemes, a 3D model with attention module, and a 3D model without attention module. We finally employ a union operation to fuse the outputs of each model to form the final segmentation. 

\subsection{Preprocessing and postprocessing}
Our preprocessing pipeline contains 5 steps: (1) non-local mean filter denoising, (2) image alignment with MNI template by rigid registration, (3) bias field correction, (4) image cropping based on region of interest (ROI), and (5) linear intensity normalization with range [0,1]. Before rigid registration, MNI template was resampled to the size of $512 \times 512 \times 128 \, \mathrm{voxel}^3$, with spatial resolution $0.377 \times 0.447 \times 1.508 \, \mathrm{mm}^3$. The ROIs are identified based on the labels of ceT1-weighted MRIs. We first make a common bounding box by taking the union of each bounding box over all labels. In order to cover all structures from all MRIs, the bounding box was then extended to the size $256 \times 128 \times 48 \, \mathrm{voxel}^3$.

In postprocessing, we extracted the largest connected component for VS from the network output. Finally, we applied the inverse transformation from the rigid registration (step 2 of preprocessing) to move the segmentations back to their original space.

\subsection{Synthesis: Image-to-image translation}
The CycleGAN \cite{zhu2017unpaired} framework is used for image-to-image translation in 2D. We first split the 3D cropped ROIs into 2D slices. Next, we feed those 2D slices to the CycleGAN for training; in this context, we consider the ceT1-weighted and T2-weighted MRIs to be two different domains. After the training process, we stack 2D slices back together to form a 3D MRI volume. The model convergence is determined based on the best performance by visual inspection at each epoch.

\subsection{Data augmentation}
Data augmentation is widely used in medical image segmentation to help minimize the gap between datasets/domains, producing more robust segmentations. Here, we design an `online' augmentation strategy during training and randomly apply data transformations to input images. These transformations are in 3 different groups:
\begin{itemize}
    \item {\textbf{Spatial augmentation.}}
3 types of random spatial augmentation are used: affine transformation with angle range of [$-10^\circ$, $10^\circ$], and scale factor from 0.9 to 1.2; elastic deformation with $\mathrm{control\ points} = 7, \mathrm{max\ displacement} = 6$; and a combination of affine and elastic deformation \cite{perez-garcia_torchio_2021}. The same spatial augmentations and parameters are applied to both MRIs and the labels.
\end{itemize}
\begin{itemize}
    \item {\textbf{Image appearance augmentation.}}
To minimize the different image appearance between MRIs from different sites and scanners, we randomly apply multi-channel Contrast Limited Adaptive Equalization (mCLAHE) and gamma correction with $\gamma$ from 0.5 to 2 to adjust image contrast.
\end{itemize}
\begin{itemize}
    \item {\textbf{Image quality augmentation.}}
In this context, image quality refers to resolution and noise level. We randomly blur the image using Gaussian kernel with $\sigma_{blur}$ from 0.5 to 1.5, we add Gaussian noise with $\sigma_{noise}= 0.01$, and we sharpen the image by $I_s = I_b + (I_b - I_{bb})\times\alpha$, where $I_s$ is sharpened image, $I_b$ is the image blurred  with a Gaussian kernel ($\sigma_{blur}$), and $I_{bb}$ is the image blurred twice with the same Gaussian kernel. In our case, we set $\alpha=10,  \sigma_{blur}=1.5$.

\end{itemize}

\begin{figure}[t]
\centering
\includegraphics[width=1\linewidth]{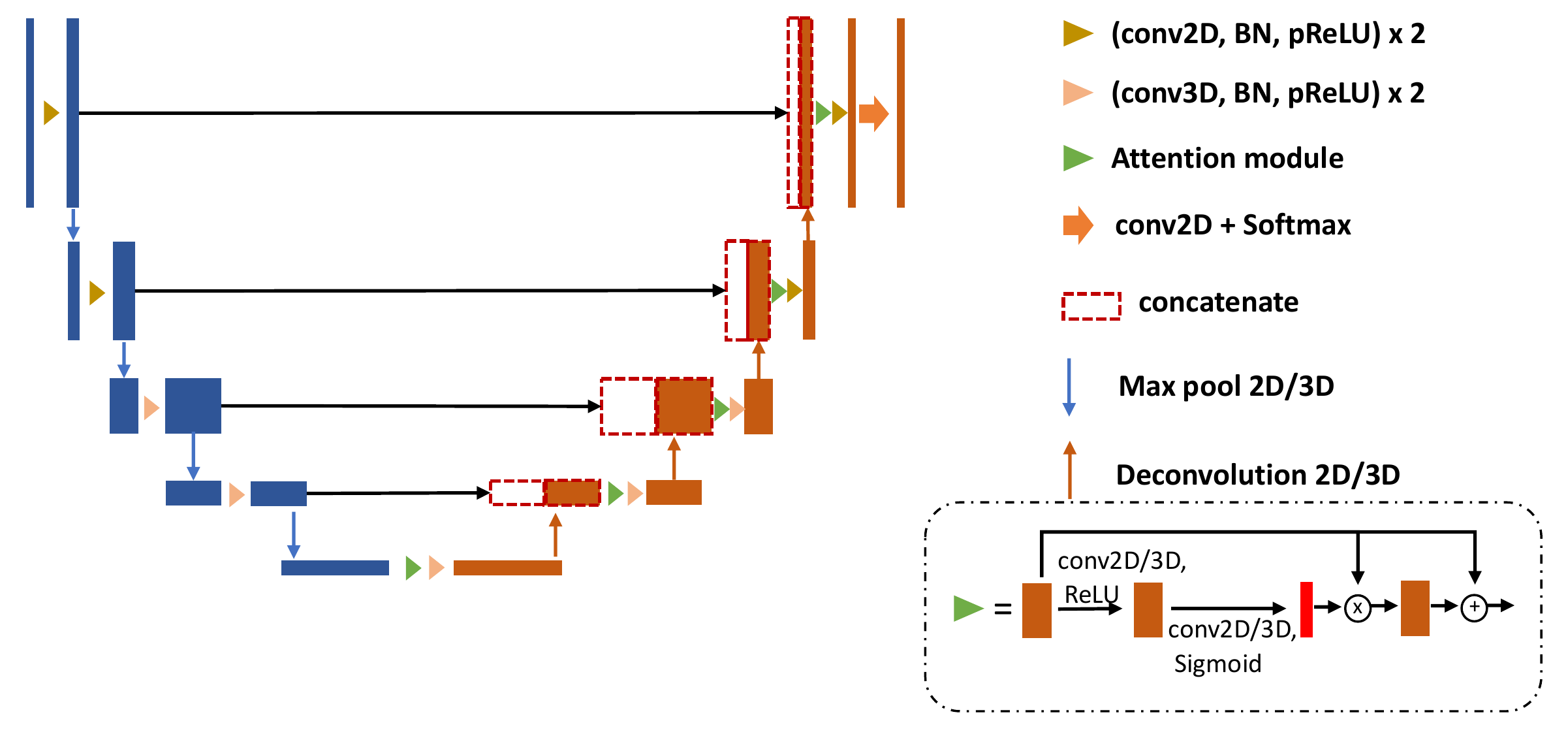}
\caption{Network architecture of 2.5D model from the work \cite{wang2019automatic,shapey2019artificial}. To deal with anisotropic image resolution, 2D convolutions are used in the first 2 levels, and 3D convolutions are used in levels 3-5.}
\label{2.5D}
\end{figure}

\subsection{Segmentation: 2.5D model and its architecture}
We leverage a 2.5D CNN \cite{wang2019automatic} model to alleviate the impact of the anisotropic image resolution. Network architecture details can be found in Fig.~\ref{2.5D}.
This 2.5D network uses both 2D and 3D convolutions to capture both in-slice and global information. 2D convolutions are used in the first 2 levels and 3D convolutions for the remainder. Adapted from U-Net \cite{ronneberger2015u}, the 2.5D model contains an encoder and a decoder, and the skip connections between encoder and decoder. The max-pooling and deconvolution operations connect the features between levels. Batch normalization and parametric rectified linear unit (pReLU) are used in the network architecture. An attention module is applied to assist in segmenting the small ROI. A second 2.5D model with identical architecture was also used in the ensemble, with the only difference being in the gamma correction parameter in the augmentation stage (with $\gamma \in [0.5, 1.5]$ rather than $[0.5, 2]$).

\begin{figure}[t]
\centering
\includegraphics[width=1\linewidth]{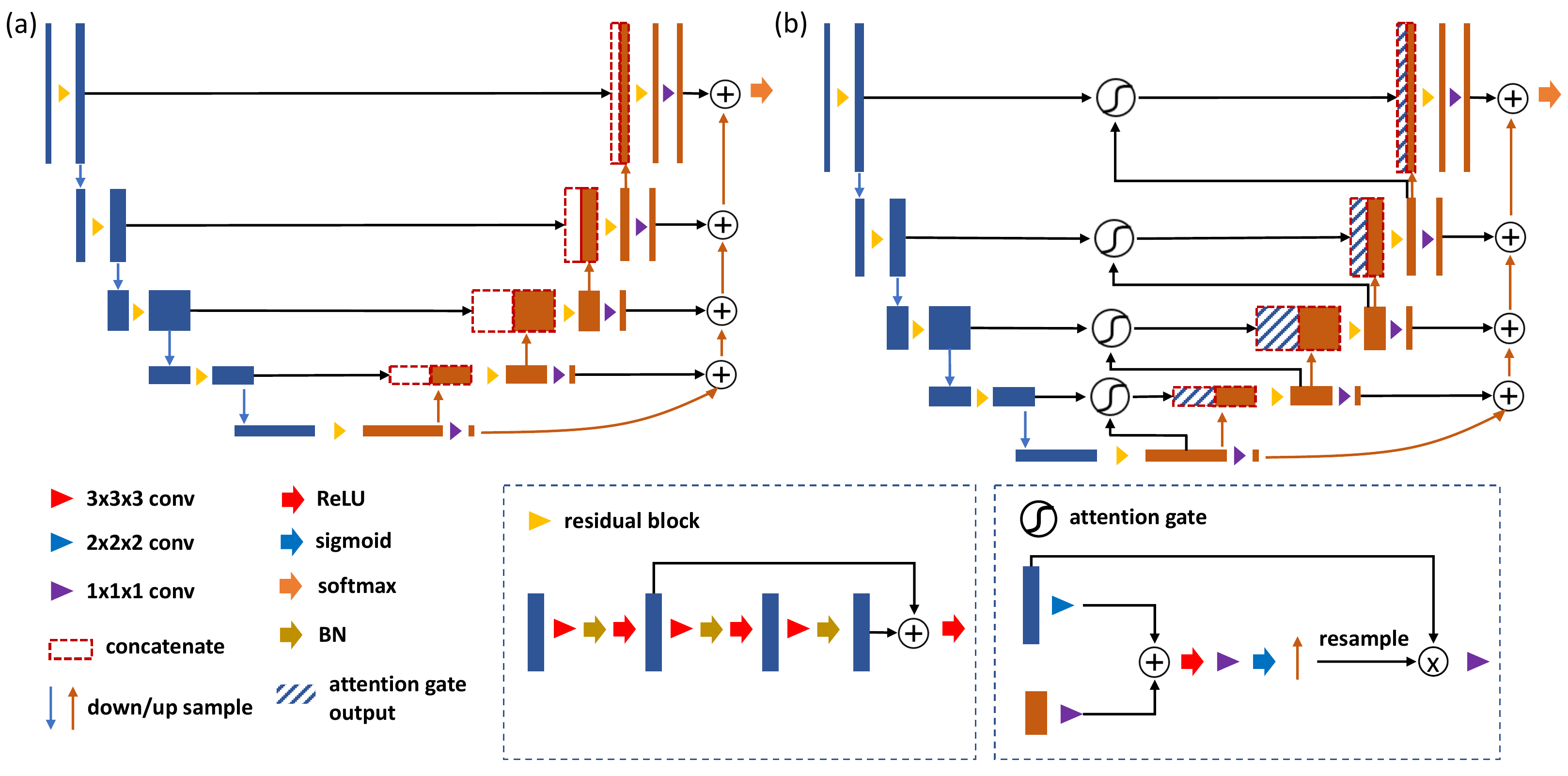}
\caption{The details of 3D models. Residual block and deep supervision are used in the 3D model. (a) 3D model without attention module, (b) 3D model with attention module.
}
\label{3D}
\end{figure}

\subsection{Segmentation: 3D model and its architecture}
We used a fully convolutional neural network for our 3D models \cite{li2021mri}. The network architecture details can be found in Fig.~\ref{3D}. Similar to a 3D U-Net, it consists of an encoder and a  decoder.  3D max-pooling and 3D nearest neighbor upsampling are used in both the encoder and decoder. We further reinforce the output by adding the feature maps at the end of each level. In one of our two 3D models, we employ an attention module in the skip connections to emphasize the small ROI and preserve the information from encoder to decoder. Note that the two 3D models are identical except the attention module.

\subsection{Implementation details}

The Adam optimizer was used with L2 penalty of 0.00001, $\beta_1=0.9$, $\beta_2=0.999$ and an initial learning rate of 0.0002 for the CycleGAN generators \cite{zhu2017unpaired} and  0.0001 for the segmentation models \cite{wang2019automatic,shapey2019artificial,li2021mri}. For the generators, the learning rate was left at the initial value for the first 100 epochs, and then dropped to 0 in following 100 epochs. For segmentation models, learning rate was decayed by a factor of 0.5 every 50 epochs. We evaluated our generator and segmentation models every epoch and selected the optimal model based on visual inspection of image quality for generator, and Dice score for segmentation. Based on the Dice loss \cite{milletari2016v}, we defined our loss function as $1-\mathrm{mean(Dice)}$ from multiple labels, with equal weight ($w_{FG}=1$) for all foreground labels and decayed weight ($w_{BG}=0.1$) for the background. The training, which has a batch size of 2, was conducted on NVIDIA GPUs and implemented using PyTorch.

\begin{table}[h]
\caption{Quantitative results in validation phase. The Dice score and average symmetric surface distance (ASSD) reported as $\mathrm{mean (stdev)}$. Baseline denotes no cropping in pre-processing, followed by synthesis and segmentation. Crop denotes the cropped input with certain size. + represents the cumulative approach based on the previous method. Bold numbers indicate the best results in each column.}
\begin{center}
    \begin{tabular}{ l  c  c  c  c }
    \hline
    \hline
    & \multicolumn{2}{c}{VS}  & \multicolumn{2}{c}{Cochlea}\\
    \hline
     Method & Dice  & ASSD & Dice  & ASSD \\ 
    \hline
    baseline & 0.408(0.284)  & 8.205(11.327) & 0.405(0.195)  & 1.697(3.684)\\
    crop ($256 \times 128 \times 96 \, voxel^3$) & 0.637(0.324)  & 2.495(7.443) & 0.603(0.219)  & 2.481(5.688)\\
    crop ($256 \times 128 \times 48 \, voxel^3$) & 0.662(0.265)  & 3.008(3.916) & 0.620(0.048)  & 0.454(0.158)\\
    + data augmentation & 0.740(0.193)  & 2.481(9.592) & 0.731(0.043)  & \bftab0.288(0.114)\\
    + ensemble (proposed) & \bftab0.794(0.156)  & \bftab0.634(0.359) & \bftab0.741(0.041)  & 0.294(0.060)\\
    \hline
  \end{tabular}
\end{center}
\label{table1}
\end{table}

\section{Results}
\subsection{Quantitative results}
Tab.~\ref{table1} displays the quantitative results of our methods in the validation phase, reported as $\mathrm{mean(stdev)}$. The initial result of our method is shown as baseline in the Tab.~\ref{table1}, which uses the resampled MRIs without any cropping as input to the 3D model with attention module. Different crop sizes in the preprocessing step leads to different results, as evidenced by the results presented in  Tab.~\ref{table1}. By adding various data augmentations, the performance of segmentation network is dramatically improved, 7.8\% and 11.1\% on Dice scores for VS and cochlea respectively. Lastly, using the model ensemble boosts the Dice score of VS to nearly 80\%. We submitted the proposed method as our final result in the validation phase of the cross-MoDA challenge.

\subsection{Qualitative results}
\begin{figure}[t]
\centering
\includegraphics[width=1\linewidth]{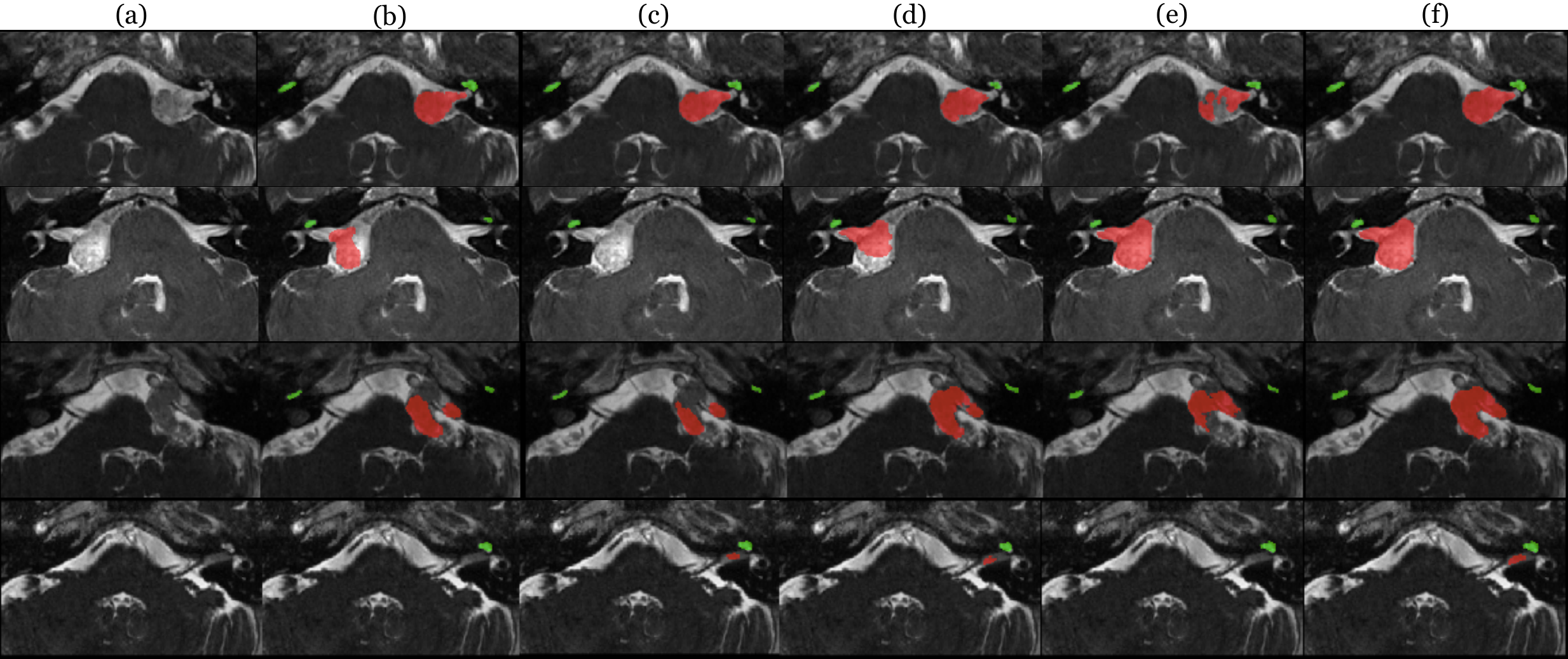}
\caption{Segmentation results from 4 different subjects. Red, VS, green, cochlea. (a) Input image. (b) 2.5D model with $\gamma \in [0.5, 2]$. (c) 2.5D model with $\gamma \in [0.5, 1.5]$. (d) 3D model without attention. (e) 3D model with attention. (f) Final segmentation. While each individual model has performance issues in some rows, the final model fusion step produces consistently good segmentations for all rows.}
\label{result}
\end{figure}

Representative qualitative results from 4 different subjects can be viewed in Fig.~\ref{result}. In the first row of Fig.~\ref{result}, we observe that our 2.5D method produces an accurate segmentation. Varying the augmentation parameters of the 2.5D model improves performance in some challenging cases  (e.g., fourth row of Fig.~\ref{result}-(c)). However, in some cases, both 2.5D models under-segments the VS, which can be seen in the second and third rows of Fig.~\ref{result}-(b,c). In such cases, 3D models compensate for this tendency of the 2.5D models. Although the attention module has good ability to capture small ROIs, which can be observed in the second row of Fig.~\ref{result}-(e), it may also lead to over-fitting (first and third rows of Fig.~\ref{result}-(e)). Thus, we also include a 3D model without attention module (first and third rows of Fig.~\ref{result}-(d)) in the model ensemble for best results. The model fusion balances the strengths of the individual models and is able to produce consistently good segmentations in a variety of images (Fig.~\ref{result}-(f)).

\section{Discussion}
There are four main design choices of our method that merit discussion: (1) cropping ROI, (2) image-to-image translation, (3) segmentation network, and (4) self-training strategy. 
\begin{itemize}
    \item {\textbf{Cropping ROI.}} Given the small size of the segmented objects and the overall MRI size, cropping an ROI is an essential preprocessing step. Using the cropped ROI as input not only reduces redundant computation and improves the computational efficiency by letting the network focus on the the  structures of interest increases accuracy. An additional benefit is the reduced GPU memory requirements. However, determining the optimal size for cropped MRIs is not straightforward. Too small ROIs might not fully contain all the structures of interest; however, too large ROIs would negatively impact the quality of the image-to-image translation and segmentation results by increasing the intensity variability in the input samples. How to balance the size of ROIs and the quality of generated MRIs from image-to-image translation remains an important problem.
\end{itemize}
\begin{itemize}
    \item {\textbf{Image-to-image translation.}} Image-to-image translation is a critical step in our pipeline and generally in UDA problems \cite{huo2018adversarial,chen2019unsupervised}. Moreover, generating well-aligned pseudo T2-weighted MRIs from source domain (i.e., from ceT1-weighted MRIs) could determine the accuracy of segmentations in the target domain, by minimizing the gap between the two domains. Thus, choosing a suitable image-to-image translation method is important. In our experiments, we employ CycleGAN \cite{zhu2017unpaired}, a popular method for unpaired image-to-image translation. However, synthesis artifacts (intensity shift between slices) appear along the depth direction of generated images in 2D CycleGAN. This is a common issue of 2D CycleGAN for translating volumetric MRIs, since slices are  mapped independently from each other during training. We also found that 3D CycleGAN requires large amount of GPU memory, and the quality of images generated by using downsampled input MRIs is not satisfactory. Furthermore, patch-based 3D CycleGAN approaches could blur MRIs along the borders of the patches. However, modifying the architecture of the generator in CycleGAN pipeline could lead to satisfactory results \cite{zhang2018translating}. Thus, improving the quality of results from image-to-image translation is another way to boost the performance. This argument is not only about the training manner of CycleGAN pipeline, but also more generally for methods such as CUT \cite{park2020contrastive} and MUNIT \cite{huang2018multimodal}.
\end{itemize}
\begin{itemize}
    \item {\textbf{Segmentation network.}}
    In our pipeline, the segmentation performance is primarily determined by the quality of the generated MRIs in the target domain. Nevertheless, given the same input MRIs, different segmentation models produce different results. In our work, we combine results from an ensemble of models to form the final segmentation, since each model could bring different advantages, even with small changes. The design choices then become the specific models and the ensembling strategy. Additional models with different hyper-parameters could thus improve the segmentation results. The good performance of the top teams in the challenge \cite{dorent2022crossmoda} suggests that the nnU-Net \cite{simpson2019large}, which is a well-known framework for biomedical image segmentation, could provide a good alternative for our ensemble models. Additionally, while we currently use a simple union operation, a more sophisticated ensembling strategy could be designed to minimize the weaknesses of each model while maximizing the strengths. 
\end{itemize}
\begin{itemize}
    \item {\textbf{Self-training strategy.}} Finally, the training strategy contributes to the accuracy of segmentation results. Compared to other teams \cite{dorent2022crossmoda}, our training strategy is potentially the biggest limitation of our work, since we just use traditional supervised learning after generating T2-weighted MRIs. Following this training strategy, we only use the generated images instead of the real images in the segmentation task, which leads to our model lacking information in the original target domain. In an ideal situation, the features of generated T2-weighted MRIs should be very close or identical to the target domain, and the segmentation model could achieve a good accuracy by leveraging those features. However, given the limitations of image-to-image translation, not all of the generated T2-weighted MRIs are well-aligned to the target domain. Self-training could be a good approach to better handle such situations in the challenge \cite{dorent2022crossmoda}; this may involve directly segmenting real T2-weighted MRIs by a pre-trained model with generated T2-weighted MRIs as inputs, and designing an algorithm to find plausible segmentations among the results. Next, we could use these plausible segmentations as `ground truths' to further fine-tune the pre-trained segmentation model. Iterating this process could further improve the segmentation accuracy. We believe that using such a self-training strategy would improve the final segmentation results. 
\end{itemize}

\section{Conclusion}
In this work, we proposed an unsupervised cross-modality domain adaptation framework for VS and cochlear segmentation. There are two parts in our framework: synthesis and segmentation. We applied various `online' data augmentations to deal with the MRIs from different sites and scanners. In addition, a model ensemble is used for increasing the performance. In the validation stage of the crossMoDA challenge, our method shows promising results.

\textbf{Acknowledgments.}
This work was supported, in part, by NIH grant R01-NS094456.

\bibliography{main} 
\bibliographystyle{splncs04}

\end{document}